\documentstyle[aps]{revtex}
\input{epsf.sty}\input{psfig.sty}
\begin{document}
\title{On the Origin of the Intensity Deficit in Neutron Compton Scattering}
\author{G.F.Reiter}
\address{Physics Department, University of Houston, Houston,  Texas, USA}
\author{P.M.  Platzman}\address{Bell Labs Lucent Technologies, Murray Hill, New Jersey,
USA}
 \maketitle
 \begin{abstract}
  Neutron Compton Scattering
measurements in a variety of materials have shown a relative
deficit in the total signal from hydrogen compared to deuterium
and heavier ions. We show here that a breakdown in the
Born-Oppenheimer approximation in the final states of the
scattering process leads to such a deficit, and may be responsible
for the effect.
\end{abstract}
\vskip 12pts PACS numbers, 61.12.Ex, 61.25.Em,82.30.Rs \vskip .5in

Deep Inelastic Neutron Scattering experiments on a variety of
systems have found the integrated intensity of the signal from
hydrogen to be smaller than expected, relative to heavy ions such
as oxygen, or even deuterium,\cite{dreis} given the well known
nuclear cross-sections for these materials. This deficit has also
been observed in electron Compton scattering.\cite{dreis3} This
has been attributed to previously unknown physics at the very
short time scales of the measurements\cite{dreis2}, quantum
entanglement between the struck proton and the other protons of
the material\cite{lk}, inadequacies in the treatment of the
data\cite{cow1,cow2}, and a failure of the usual van Hove
scattering formula to describe the data\cite{gid}. We show here
that the effect can be explained within the usual formalism by a
breakdown of the Born Oppenheimer approximation in the final state
of the scattering process, which contains a very rapidly moving
proton with sufficient energy to mix the electronic states of the
system. These corrections are present for any scattering
experiment, but only become significant for high momentum
transfer.

We begin with the usual expression for the scattering
cross-section when  the spin of the neutron is not observed, in
the large q limit in which  only incoherent scattering remains.
\begin{equation}
S(\vec{q},\omega)
=\Sigma_{\vec{n}}|<0|e^{-i\vec{q}.\vec{R_1}}|E_{\vec{n}}>|^2\delta(\omega-{(E_{\vec{n}}-E_0)\over
\hbar} ) \label{sw}
\end{equation}
where the ground state is taken to be the Born Oppenheimer ground
state,
\begin{equation}
<\vec{r},\vec{R}|0>=\Phi_{0,0}(\vec{R})\alpha_0(\vec{r},\vec{R})
\end{equation}
 $\vec{r}$ are the coordinates of the electrons and $\vec{R}$ the
coordinates of the ions. The $|E_{\vec{n}}>$ are energy
eigenstates of the complete system of electrons and ions.
$\alpha_0(\vec{r},\vec{R})$ is the solution of the electronic
problem with the protons and other ions at the positions
$\vec{R}$,
 and $\Phi_{0,0}(\vec{R})$ is the  ground state for the heavy particles
 in the potential energy surface defined by $|\alpha_0>$. $\vec{R_1}$
 is the component of $\vec{R}$ denoting the position of the struck particle, There are in addition, a complete set
of excited electronic states, for the ions at the same positions,
denoted by $\alpha_i(\vec{r},\vec{R})$. It is evident from Eq.
\ref{sw}that
\begin{equation}
\int S(\vec{q},\omega)d\omega=1
\end{equation}
for any q. The total intensity of the incoherent scattering is
proportional  to this integral times the total scattering
cross-section for the  scattering particle. Our point of view is
that if this appears not to be the case, then some intensity is
not being accounted for in the experimental integration.

 It is usually assumed at this
point, that the energy eigenstates of the full Hamiltonian can be
adequately represented by the ground state for the electronic
system and the excited states for the ions in the Born Oppenheimer
potential for that ground state. The energies $E_{\vec{n}}$ are
the sum of the electronic energy, $E_n^{'} (\vec{R})$ and the
energy of the heavy ions in the effective potential defined by
this energy surface. However, we wish to consider the limit that
the transferred momentum is large, so that the recoiling proton
has sufficient energy to mix  the electronic states. In short, we
must go beyond the Born Oppenheimer approximation in expanding the
state $e^{i\vec{q}.\vec{R_1}}|0>$ in energy eigenstates, and
consider the full Hamiltonian for the system, which includes the
kinetic energy of the  struck proton. We will treat this kinetic
energy as a perturbation, to the extent that it acts on the
electronic wavefunctions, and first calculate the energy
eigenstates to lowest order. We take the full Hamiltonian to be
\begin{equation}
H=H_0(\vec{r},\vec{R})+ \Sigma_i{P_i^2\over{2M}}
\end{equation}
  The BO  energy eigenstates
,$|\Phi_{n,j}>|\alpha_n >$satisfy
\begin{equation}
H_0(\vec{r},\vec{R})\alpha_n(\vec{r},\vec{R})=E_n^{'}(\vec{R})\alpha_n(\vec{r},\vec{R})
\label{bo}
\end{equation}
and
\begin{equation}
(\Sigma_i{P_i^2\over{2M}}+E_n^{'}
(\vec{R}))\Phi_{n,j}(\vec{R})=E_{n,j}\Phi_{n,j}(\vec{R})
\end{equation}
 In addition to the kinetic energies of the heavy particles,
obtained from $(\Sigma_i{P_i^2\over{2M}}|\Phi>)|\alpha_n>$ the
leading term in the perturbation when acting on a state of the
form $|\Phi>|\alpha_n>$ is
\begin{equation}
{\Sigma}_{i} \vec{P_{i}}|\Phi >.\vec{P_{i}}|\alpha_{n}>/M
\end{equation}
The action of the momentum of the heavy particle on the electronic
states is,  from Eq. \ref{bo}
\begin{equation}
<\alpha_m |\vec{P_i }|\alpha_n>=i\hbar<\alpha_m |{{{{\delta
H_0}\over{\delta \vec{R_i}}}} |\alpha_n>/(E_n^{'}-E_m^{'})}
\end{equation}

where the matrix elements are taken over the electronic
coordinates, and the positions of the heavy particles are the same
for all the states. This perturbation mixes the original
Born-Oppenheimer energy eigenstates. For all the heavy particles
except the one struck by the neutron, the momentum operator acting
on the wavefunction of the particle gives a small overall
contribution, which is the basis for the usual Born-Oppenheimer
approximation. However for the struck particle, particle 1 in Eq.
\ref{sw}, this operator gives a large term, of order $\hbar
\vec{q}$, and cannot be neglected.

  We have then that the energy
eigenstates for the full Hamiltonian are given to first order by
\begin{equation}
|E_{n,i}>=|\Phi_{n,i}>|\alpha_n>+\Sigma_{m,j}<\Phi_{m,j}|<\alpha_m|{{i\hbar{{\delta
H_0}\over{\delta
\vec{R_1}}}}\over{(E_{n,i}-E_{m,j})(E_n^{'}-E_m^{'})}}|\alpha_n>\cdot{\vec{P_1}\over
M} |\Phi_{n,i}>|\Phi_{m,j}>|\alpha_m>
  \label{bo2}
\end{equation}
We will neglect for the moment the shift of the energies for these
states, as it is not essential for this work. The important point,
is that because of the presence of the fast proton, the excited
states have a significant overlap with the ground state, and as a
consequence, there is some amplitude for the scattered neutron to
transfer some of its energy to the electronic system.  When Eq.
\ref{bo2} is used in Eq. \ref{sw}, we see that the only component
of the excited states wave function that is significant is that
which couples it to the ground state. The sum in Eq. \ref{sw}
extends over the indices of the electronic states, and for each
electronic state, one sums over the spectrum of eigenstates of the
ionic system in that electronic state. We have then that there is
a contribution to the sum in Eq. \ref{sw} from each of the the
excited electronic states, $S_n(\vec{q},\omega)$ of the form

\begin{equation}
S_n(\vec{q},\omega)=\Sigma_{i,j}|<\Phi_{0,0}|e^{-i\vec{q}.\vec{R_1}}|\Phi_{0,j}><\Phi_{0,j}|{{\hbar}\over{M}}.<\alpha_0|{{{{\delta
H_0}\over{\delta
\vec{R_1}}}}\over{(E_{n,i}-E_{0,j})(E_n^{'}-E_0^{'})}}|\alpha_n>\cdot\vec{P_1}|\Phi_{n,i}>|^2\delta(\omega-
{(E_{n,i}-E_{0,0})\over\hbar}) \label{sp}
\end{equation}

 We are going to approximate the electronic matrix element in Eq.
  \ref{sp} by evaluating it at the equilibrium positions of the
 heavy  particles.
 Since only ionic states for which the struck particle has a momentum in
the vicinity of $\hbar q$ will have any significant overlap with
the state $<\Phi_{0,0}|e^{-i\vec{q}.\vec{R_1}}$, we can assume the
$|\Phi_{0,j}>$ that are important in the sum have this property.
Then this must be true of the states $|\Phi_{n,i}>$ as well.
Furthermore, for the heavy particles which have not been struck,
henceforth referred to as the bath, the wavefunctions and energies
in both the ground and excited states are unaffected, initially,
by the scattering event.  Hence the change in the heavy particle
energies in both the excited and ground electronic states is
dominated by the kinetic energy of the struck particle, ${(\hbar
q)^2\over{2M}}$. Thus we can replace $(E_{n,i}-E_{0,j})$ in the
denominator of Eq. \ref{sp} by $(E_n^{'}-E_0^{'})$. The electronic
matrix element is then independent of the heavy particle energy
indices, and we can use any basis for the particle wavefunctions
that we like. In particular, we choose a plane wave basis. This
also allows us to evaluate the action of $\vec{P_1}$ , which we
will replace with $\hbar q$. This approximation neglects the
momentum of the particles in the initial state compared to $\hbar
q$ when evaluating the matrix elements. In the energy delta
function, we will keep the full kinetic energy ${(\hbar
(\vec{p}+\vec{q}))^2\over{2M}}$  in the final state, and replace
$E_{0,0}$ by $E_0^{'}+{p^2\over{2M}}$. None of these
approximations are essential for our main point, but serve to
simplify the result. We find then that the excited electronic
states contribute to $S(\vec{q},\omega)$

\begin{equation}
\Sigma_n S_n(\vec{q},\omega)={\Sigma}_{n}
|{{\hbar^2\vec{q}}\over{M}}.<\alpha_0|{{{{\delta H_0}\over{\delta
\vec{R_1}}}}\over{(E_n^{'}-E_0^{'})^2}}|\alpha_n>|^2\int
n(\vec{p})\delta(\omega-(E_n^{'}-E_0^{'})/\hbar-{{\hbar
q^2}\over{2M}}-{{\vec{p}.\vec {q}}\over{ M}}) d\vec{p}
\label{swex}
\end{equation}

where $n(\vec{p})$ is the diagonal matric element of the one
particle  density matrix for particle 1 in the momentum
representation. We see that these terms give the same functional
form in q and $\omega$ as the usual impulse approximation result
but shifted to high energies by the difference in energies of the
electronic states.

The total intensity in these terms is compensated for in the sum
in Eq.  \ref{sw} by the reduction in intensity of the scattering
in which there is no change in the electronic wave function.  The
ground state of the electronic system, in the presence of the fast
proton, is no longer $|\alpha_0>$, but is given by Eq. \ref{bo2}.
When used in the sum in Eq. \ref{sw}, it must be normalized. The
normalization of this state leads to a reduction of intensity of
the primary signal, again with the same approximations, by an
amount $\Delta I$
\begin{equation}
\Delta I=\Sigma_n|{{\hbar^2\vec{q}}\over{M}}.<\alpha_0|{{{{\delta
H_0}\over{\delta
\vec{R_1}}}}\over{(E_n^{'}-E_0^{'})^2}}|\alpha_n>|^2 \label{res1}
\end{equation}
This just compensates for the intensity transferred to the excited
electronic states.  To the extent that one measures only the
intensity in the primary peak, Eq. \ref{res1} gives the apparent
deficit in the crossection of the struck particle. We see that
this varies with $q^2$ and $1/M^2$, as well as the particulars of
the electronic system.

Although a detailed calculation is necessary to evaluate the
magnitude of the deficit in the intensity of the main peak, we can
get some idea of the size of the effect  by introducing
dimensionless variables. A characteristic length scale for
molecular systems is the radius of the Bohr orbit, $a$. A
characteristic energy scale is ${e^2}\over 2a$. Since
\begin{equation}
{{{{\delta H_0}\over{\delta \vec{R_1}}}}=\Sigma_i
{{e^2}\over{|\vec{r_i}-\vec{R_1}|^2}}} \label{pot}
\end{equation}
where the sum in Eq.\ref{pot} is over all the electrons in the
system, we find that an estimate of the  magnitude of the effect
is given by
\begin{equation}
\Delta I \approx ({{{2
N_{eff}\hbar^2q/a}\over{M}}\over{e^2/2a}})^2 \label{sc}
\end{equation}

 where $N_{eff}$ is the effective number of terms in
the sum appearing in Eq. \ref{pot}.

 For a
q of 50$\AA^{-1}$  and $N_{eff}$ of 3, $\Delta I\approx .15$.
Using a perhaps more realistic scale for water, which would be the
energy gap of about 6 ev, we would have an increase by a factor of
four. The result is also very sensitive to the value of Neff. It
is clear that without evaluating the integrals, there is
considerable uncertainty in the magnitude of the effect, but that
nevertheless, it is potentially of the right order of magnitude to
explain the deficits.  However, the result above cannot explain
two features of the phenomenology of the deficit as it is observed
in a variety of systems. The deficit is not  observed to vary with
$q^2$\cite{dreis2,jert} in the Neutron Compton Scattering
measurements on water. Although the results of the Electron
Compton Scattering experiments in formvar\cite{dreis3} are
consistent with a $q^2$ dependence of the deficit for the small
range of q measured, the great bulk of the available data, over a
much wider range in q, is obtained using neutrons. The data on the
deficit in Nb and Pd hydrides \cite{kl}does show a strong q
dependence, which, however saturates at large q. The saturation
value of $\Delta I$ is about .4, which is four times the value in
the deuterated Nb hydride, as the theory above would predict, but
the deuterated material shows no variation with q.

 Also,  as it stands, there is no explanation for the
dependence of the deficit on the relative concentration of
deuterium and ordinary water.\cite{dreis} Since the matrix
elements and the density of electronic states that determine the
amplitude of the effect are presumably nearly the same in $D_2O$
and $H_2O$, we would expect no concentration dependence from the
results above. Both of these inadequacies point to the coupling of
the electronic states in the presence of a fast proton being too
strong to use the weak coupling perturbation theory presented so
far. The effect of the higher order terms in the coupling may be
accounted for by introducing linewidths for the excited states due
to transfer of energy in the electronic system to the ions.
 These linewidths are strongly enhanced by the presence of the fast
proton.

The time scale for the relaxation of the density correlation
function, $({q\Delta p\over M})^{-1}$\cite{rs}, where $\Delta p$
is the momentum distribution width, is about 1 femtosecond in
water, for typical momentum transfers of 35$\AA^{-1}$. This
corresponds to a momentum width for the protons of about
$5\AA^{-1}$\cite{ramp}. It is known that the relaxation time for
the lowest excited electronic state in water to the so called
solvated electron state, is about 150 femtoseconds,\cite{rossky}
considerably longer than the time scale we are interested in.
However, that is when the protons are moving at velocities
comparable to ${\Delta p}\over M$ above. When one of these protons
has momentum of 30-40 $\AA^{-1}$ the relaxation rate could be a
factor of fifty faster, since this rate has a component that
varies as $q^2$. That rate is comparable to the typical time scale
of the scattering, and the decay of the excited state can play a
major role in determining the intensity deficit. Furthermore, it
is known\cite{bit} that the excited electron wave function extends
over 5-10 water molecules, and that the rate of transfer of energy
to the protons or deuterons depends on a decoherence time scale
that is longer by a factor of two in $D_2O$ than in  $H_2O$,
providing a mechanism for the concentration dependence of the
deficit. Although we are not be able to compute the electronic
properties needed to make this argument precise, it is plausible
that a stong coupling theory that includes the decay of the
excited states can account for the concentration dependence in
water.

To go beyond lowest order perturbation theory, we will use a mode
coupling approximation to include the decay of the excited states.
The intermediate scattering function, S(q,t), is

\begin{equation}
S(\vec{q},t)=<0|e^{-i\vec{q}.\vec{R_1}}e^{i\vec{q}.\vec{R_1(t)}}|0>=<0|e^{-i\vec{q}.\vec{R_1}}e^{{i\over\hbar}(H_0+\Sigma{{P_i}^2\over
2m})t} e^{i\vec{q}.\vec{R_1}}e^{-{i\over\hbar}(H_0+{\Sigma
{P_i}^2\over 2m})t}|0>
 \label{lap}
\end{equation}
Since
\begin{equation}
\Sigma{{p_i}^2\over
2M}\Phi_{0,0}(\vec{R})=(E_{0,0}-E_0^{'}(R))\Phi_{0,0}(\vec{R})
\end{equation}
it follows that the action of the Hamiltonian on the ground state
in Eq. \ref{lap} can be replaced by

\begin{equation}
e^{-{i\over\hbar}E_{0,0}t}\Phi_{0,0}(\vec{R})=e^{-{i\over\hbar}E_0^{'}(R)t}e^{{-{i\over\hbar}\Sigma{
P_i^2\over 2M}t}}\Phi_{0,0}(\vec{R})
\end{equation}
Having done this, the function $e^{-{i\over\hbar}E_0^{'}(R)t}$
commutes with the density operator, $e^{i\vec{q}.\vec{R_1}}$, and
can be combined with the Hamiltonian acting on the final state.
There is an additional term, arising from the  commutator of the
kinetic energy term in Eq. \ref{lap} with $E_0^{'}(R)$. This gives
the effects of the forces acting on the heavy particles. This term
has no direct effect on the impulse approximation limit,
\cite{rs}, $q\rightarrow\infty, t\rightarrow 0$, qt finite, as it
is  of order $qt^2$. It leads to
 traditional final state corrections of order 1/q that are included in the
 data analysis.  Thus we can include the ground
state electronic energy in the exponent in Eq. \ref{lap}. The
kinetic energy terms that do not involve the struck particle can
also be approximated as commuting with the electronic energy,
since this is the essence of the Born-Oppenheimer approximation.
The noncommutative terms are responsible for the thermalization of
the electronic states. These lead to the relaxation processes
mentioned earlier, and  occur on a much longer time scale than the
one we are interested in. We will therefore neglect the commutator
of the kinetic energy of the unstruck particles with the
electronic Hamiltonian. We can then commute the kinetic energy for
the unstruck particles through the density operator and eliminate
it from the direct time evolution of the density operator.

Finally, we can approximate the operator ${P_1}^2$ by
\begin{equation}
{P_1}^2={P_1^1}^2+2P_1^1.P_1^2 \label{p12}
\end{equation}

where the operator $P_1^1$ acts only on the proton wave function,
and the operator $P_1^2$ acts only on the parameters $\vec{R}$ of
the electronic wavefunction. In its action on the ground state, we
will ignore the second term on the right in Eq. \ref{p12}, which
is, again, the Born Oppenheimer approximation.

 We will also replace
$P_1^1$ by $\hbar \vec{q}$ in the second term of Eq. \ref{p12},
when this operator  acts on final states for which the momentum is
approximately  $\hbar \vec{q}$.   The full Hamiltonian acting on
the final state is now the sum of two commuting terms. allowing
the operator ${P_1^1}^2$ to act on $e^{i\vec{q}.\vec{R_1}}$  we
can then express the intermediate scattering function as
\begin{equation}
S(\vec{q},t)=<0|e^{-i\vec{q}.\vec{R_1}}e^{{i\over\hbar}(H_0-E_0^{'}+{\hbar
\vec{q}.\vec{P_1^2}\over M})t} e^{i\vec{q}.(\vec{R_1}+{\vec{P_1^1
}\over M }t)}|0> \label{sqt1}
\end{equation}

If we neglect the perturbation ${\hbar \vec{q}.\vec{P_1^2}\over
M}$ in Eq. \ref{sqt1} we obtain immediately the impulse
approximation result for S(q,t), making clear again that it is the
mixing of the Born-Oppenheimer eigenstates that is the source of
the intensity deficit.

 Eq. \ref{sqt1} can be expressed in more
detail, making use of the identity $ e^{{i\hbar q^2\over
2m}t}e^{i\vec{q}.\vec{R_1}}e^{i\vec{q}.({{\vec{P_1^1}\over M
}})t}=e^{i\vec{q}.(\vec{R_1}+{\vec{P_1^1}\over M })t}$, as
\begin{equation}
S(\vec{q},t)=e^{{i\hbar q^2\over 2M}t} \int
d\vec{r}d\vec{R}\Phi_{0,0}^*(\vec{R})\alpha_0^*(\vec{r},\vec{R})e^{{i\over\hbar}(H_0-E_0^{'}(\vec{R})+{\hbar
\vec{q}.\vec{P_1^2 }\over M})t}e^{{i{\vec{q}.\vec{P_1^1}\over
M}t}} {\alpha}_0(\vec{r},\vec{R})\Phi_{0,0}(\vec{R}) \label{sqt}
\end{equation}

For each value of R, we see that we now have a standard
perturbation problem for the action of the Hamiltonian on the
electronic wave function.

It is clear from the above, that it is the evolution of the
electronic state in the presence of the fast proton that must be
calculated to obtain the deficit. We choose to do the perturbation
calculation for the electronic propagator in the frequency domain
rather than directly in the time domain.  Using standard
projection operator methods for the the resolvent operator for the
electronic Hamiltonian we can show

\begin{equation}
<\alpha_0|[z-H_0+E_0^{'}(\vec{R})+{\hbar \vec{q}.\vec{P_1^2}\over
M}]^{-1}|\alpha_0>=[z-\Gamma_0(z)]^{-1} \label{res2}
\end{equation}
where
\begin{equation}
{\Gamma}_0(z) ={\Sigma}_{n,m}^{'}
({{{\hbar^2\vec{q}}\over{M}}})^2<{\alpha}_0|{{{{\delta
H_0}\over{\delta
\vec{R_1}}}}\over{(E_n^{'}-E_0^{'})}}|{\alpha}_n><{\alpha}_n| [z
-(H_0-E_0^{'})/\hbar-{{\hbar \vec{q}.\vec{P_1^2 }}\over
M}]^{-1}|{\alpha}_m><{\alpha}_m|{{{{\delta H_0}\over{\delta
\vec{R_1}}}}\over{(E_m^{'}-E_0^{'})}}]^{-1}|{\alpha_0>}
\label{res3}
\end{equation}

 The prime on the summation in Eq, \ref{res3} indicates the that the the
ground state is omitted from both sums. Furthermore, the
perturbation ${\hbar \vec{q}.\vec{P_1^2 }\over M}$   operator
should be understood as having no matrix elements with the ground
state.

We will approximate the sum in Eq. \ref{res3} by keeping only the
diagonal terms. These are the leading terms in the perturbation
series in powers of q. In this case, we have
\begin{equation}
{\Gamma}_0(z)={\Sigma}_n^{'}{{
|{{\hbar^2\vec{q}}\over{M}}.<\alpha_0|{{{{\delta H_0}\over{\delta
\vec{R_1}}}}\over{(E_n^{'}-E_0^{'})}}|\alpha_n>|^2\over {z
-(E_n^{'}-E_0^{'})/\hbar-\Gamma_n(z)}}} \label{res}
\end{equation}
where
\begin{equation}
{\Gamma}_n(z)={\Sigma}_m^{'}{
|{{\hbar^2\vec{q}}\over{M}}.<\alpha_n|{{{{\delta H_0}\over{\delta
\vec{R_1}}}}\over{(E_m^{'}-E_n^{'})}}|\alpha_m>|^2\over {z
-(E_m^{'}-E_0^{'})/\hbar-\Gamma_m(z)}} \label{gam}
\end{equation}
The prime on the sum in Eq. \ref{res} indicates that the ground
state is omitted, and in Eq. \ref{gam} that both the ground state
and the state m=n are omitted.  Eqs. \ref{res},\ref{gam} are an
approximation, which should be understood as self-consistently
defining the self energies. We will not attempt to calculate
these. For our purposes, it is sufficient to note that  these self
energies are proportional to $q^2$ for small q, and can become
independent of q for large q.

 The energies in Eq.\ref{res} depend on the position of the proton,
which is to be averaged over in Eq. \ref{sqt}. The effect of this
averaging is to produce a distribution of energies in the
  resolvent expression Eq. \ref{res}  that is
independent of q and will therefore be neglected in comparison
with the self energy of the excited states. However, while
formally smaller in powers of q, the electronic energies are large
compared to ${\hbar q\Delta p\over M}$, and their variation can
introduce a time scale comparable to the decay time of
$S(\vec{q},t)$ for moderate values of q. In the work of Schwartz
et al\cite{bit}, it is shown that the decoherence factor
$<e^{{i\over\hbar}(E_n^{'}(\vec{R})-E_0^{'}(\vec{R}))t}>$ that
arises from this averaging has a decay time of about 5
femtoseconds for deuterium, and about half that for
protons.\cite{bit} Therefore the numerical value for $\Gamma_0(z)$
can be strongly influenced by this averaging, which must also
include, in a realistic calculation, the average over spatial
configurations of surrounding molecules. We believe this
difference in decoherence time is responsible for the
concentration dependence of the deficit in $H_{2}0$-$D_{2}0$
mixtures. The shorter decoherence time produces smaller relaxation
rates, smaller values for $\Gamma_0(z)$ and hence larger deficits.
We neglect it here for simplicity only, in order to focus on the
main point, the qualitative behavior of the intensity deficit with
q, and evaluate the energies at the equilibrium position of the
proton. The electronic propagator(resolvent) then becomes
independent of $\vec{R_1}$ and the momentum of the proton is
conserved by the dynamics.  We find then, that the Laplace
Transform of Eq. \ref{sqt1} is
\begin{equation}
\int_0^\infty dte^{-izt}S(\vec{q},t)dt= -i\int
d\vec{p}n(p)[z-{{\hbar q^2}\over{2M}}-{{\vec{p}.\vec {q}}\over{
M}}-\Sigma_{n}{ |{{\hbar^2\vec{q}}\over{M}}.<\alpha_0|{{{{\delta
H_0}\over{\delta
\vec{R_1}}}}\over{(E_n^{'}-E_0^{'})}}|\alpha_n>|^2\over {z
-(E_n^{'}-E_0^{'})/\hbar-{{\hbar q^2}\over{2M}}-{{\vec{p}.\vec
{q}}\over{ M}}-\Gamma_n(z-{{\hbar q^2}\over{2M}}-{{\vec{p}.\vec
{q}}\over{ M}})}}]^{-1} \label{ans}
\end{equation}

We can recover the results of the previous weak coupling analysis
by ignoring $\Gamma_n(z )$ in the denominator of the expression in
the sum in  Eq. \ref{ans}. The neutron scattering function,
$S(\vec{q}, \omega)$ in Eq. \ref{sw}  is obtained by letting
z$\rightarrow\omega-i\epsilon$ in Eq. \ref{ans} and taking the
real part of the resultant expression.  The amplitude of the pole
at $\omega={{\hbar q^2}\over{2M}}+{{\vec{p}.\vec {q}}\over{ M}}$
is $1/(1-{\delta \Gamma_0(z)\over \delta z}|_{z=\omega})$ where
$\Gamma_0(z)$ is  the sum in Eq.\ref{ans}. This  is the same
result as Eq.\ref{res1}, to lowest order in q. With the inclusion
of linewidths for the excited states, and hence lineshifts as
well, we have the possibility of this intensity deficit saturating
as q$\rightarrow\infty$. What is required is that the lineshifts
become comparable to the energy differences from the ground state
for the excited states for which the matrix elements in the
numerator are significant. While we cannot show that this occurs
for the values of q in the experiments, since we cannot evaluate
the electronic matrix elements, we can show that in the limit of
very strong coupling, in which the perturbation matrix elements
are comparable or greater than the energy level separations, such
a result holds. In Eq.\ref{sqt}, we can write
\begin{equation}
e^{{i\over\hbar}(H_0-E_0^{'}(\vec{R})+{\hbar \vec{q}.\vec{p_1^2
}\over M})t}\approx e^{{i\over\hbar}(H_0-E_0^{'}(\vec{R}))t}e^{{i{
\vec{q}\cdot\vec{P_1^2 }t}\over M}}e^{{i\over 2}{\vec{q}\over
M}\cdot{{{{\delta H_0}\over{\delta \vec{R_i}}}}t^2}}\label{bch}
\end{equation}
if we neglect higher order terms in the Baker-Campbell-Hausdorf
formula. The last term in Eq. \ref{bch} is of order $qt^2$. While
this is formally of higher order than the terms kept in the
impulse approximation limit, the coefficient is determined by the
forces on the proton in the excited states, and is on the scale of
electronic energies. For times of the order of the characteristic
times that are actually achieved in the experiments, the last
phase factor in Eq. \ref{did} is approximately

\begin{equation}
{{e^2\over 2a^2}({q\Delta p\over M})^{-1}\over \Delta p}\approx 9
\end{equation}

and cannot be neglected. It has been included in the discussion
above. It is nevertheless useful to consider the very large q
limit, where this phase factor can be neglected.   We then find
that S(q,t) can be written simply as
\begin{equation}
S(\vec{q},t)\approx e^{{i\hbar q^2\over 2M}t}\int
d\vec{R}\Phi^*_0(\vec{R})\Phi_0(\vec{R}+{{\hbar\vec{q_1}t}\over
M})<\alpha_0(\vec{R})|\alpha_0(\vec{R}+{{\hbar\vec{q_1}t}\over
M})>\label{did}
\end{equation}
The notation $\vec{q_1}$ in Eq. \ref{did} indicates that it is a
vector corresponding to a displacement of the coordinate of
particle 1. It should be noted that the electronic matrix element
is independent of $\vec{R}$ at t=0, and will be weakly dependent
on $\vec{R}$ for subsequent times, while the nuclear wavefunction
depends strongly on $\vec{R_1}$ Ignoring the position dependence
of the electronic overlap, we can then write the nuclear overlap
integral in the momentum representation, and obtain the usual
impulse approximation result for $S(\vec{q},t)$ multipled by a
time dependent factor. Our point here is that this factor
approaches a constant in the large q limit for times of the order
of the characteristic time. Since the time for which
$S(\vec{q},t)$ is significant is $({q\Delta p\over M})^{-1}$, the
electronic overlap integral, which is responsible for the
intensity deficit, involves the two states separated by a finite
amount,${\hbar\over\Delta p}$. Or using the uncertainty relation,
by a separation comparable to the localization distance of the
proton in the ground state. The reduction in intensity, therefore,
is independent of q in the very strong coupling limit, and
proportional to $q^2$ in the weak coupling limit. While we are not
in the very strong coupling limit, we think it plausible that the
experiments are being done in an intermediate region, with the
perturbation comparable to energy level differences, and showing a
weak q dependence. This could lead to a variety of dependences of
the deficit on the electronic properties, as observed.  While most
experiments show an increase in the deficit with q, for instance,
recent experiments on LaH2 and LaH3 show  a slight decrease with
increasing q at large q.\cite{abdul}

If the intensity is transferred, it is important to know why it
has not so far been observed. In water, the minimum excitation
energy is about 6 ev, so that the intensity that is shifted by
exciting the electronic system would be on the high energy side of
the main signal by at least that amount. It would continue to
higher energies by an amount that depended upon the sum of the
widths of the occupied and unoccupied levels that were involved in
the transitions. Using recent calculations of Cabral do Couto et
al\cite{cc}, we find this to be on the order of 30ev if the
dominant contribution to the sums in Eq.\ref{res} come from the
valence and conduction band, but the widths could be much larger
if  there is significant excitation of higher and/or deeper bands.
 It is difficult to say where the center of gravity
of the shifted spectral density would lie, but it is easily
conceivable that it lies at energies that correspond to transit
times at Vesuvio of less than 50 milliseconds, which is where the
data for water stops.  It is also the case, that because of the
collapse of the phase space for short times, where a large range
of momentum transfers are incorporated in a small change in the
time of flight, the signal may be missed due to detector
saturation.  The data for LaH2 and LaH3 is more problematical, as
there is a band gap of about .5 ev in LaH3, while LaH2 is a
metal\cite{gup} and yet there is very little difference in the
magnitude of the intensity deficit. This could be explained by the
center of gravity of the absorbtion being at energies much higher
than the gap.

It is evident that to give a quantitative explanation for the
intensity deficit in various materials, it is necessary to be able
to calculate the integrals over the electronic levels with some
accuracy. We have not done this, and indeed, it is possible that
the effects we are calculating may be too small to explain the
observations. What we have shown is that a straightforward
physical effect, the mixing of Born-Oppenheimer levels  and an
increase in the linewidth   of excited electronic levels when a
proton of large momentum is present, effects  that must be
operative at some level in any case, have the capacity to account
for the phenomenology of the intensity deficits in Neutron Compton
Scattering, and we have provided a framework in which to
accurately calculate these effects.

\begin{acknowledgements}
 We would like to thank Eric Bittner and  Sylvio Canuto for
several useful conversations and, P. Cabral do Couto, Jerry Mayers
and Tyno Abdul-Redah for making available work prior to
publication. The authors are supported by DOE Grant 1-5-555229
\end{acknowledgements}


\begin{references}

\bibitem{dreis} C. A. Chatzidimitriou-Dreismann, T. Abdul-Redah, F. Streffer, J. Mayers, Phys. Rev. Letts. {\bf 79}, 2839,(1997);
\bibitem{dreis3}C. A. Chatzidimitriou-Dreismann, M. Vos, C. Kleiner, T.
Abdul-Redah, Phys. Rev. Letts.  {\bf 91}, 057403,(2003);
\bibitem{dreis2}C. A.
Chatzidimitriou-Dreismann, T. Abdul-Redah,F. Streffer, J. Mayers
J. Chem. Phys. {\bf 116}, 1511,(2002);
\bibitem{lk}E. B. Karlsson and S. W. Lovesey, Phys. Rev. A
61, 062714,(2000);E.B. Karlsson, Phys. Rev. Letts. {\bf 90},
095301,(2003)
\bibitem{cow1}J.J. Blostein, J. Dawidovski and J.R. Granada, Physica B {\bf 304}, 357,
(2001)
\bibitem{cow2}  R. Cowley, J. Cond. Matt. Phys. {\bf 15}, 4143, (2003)
\bibitem{gid}N. Gidopoulis, (preprint)
\bibitem{rs}R. Silver and G.reiter, Phys.Rev.Letts.{\bf
54},1047,(1985)
\bibitem{jert}T. Abdul-Redah and J. Mayers, "Neutron Inelastic Cross-Sections at eV Energy Transfer",
(submitted
to J. Cond. Matt. Phys. )
\bibitem{kl} E. Karlsson et al, Phys. Rev. B, {\bf 67}, 184108,
(2003)
\bibitem{ramp}G. Reiter, J.C. Li, J. Mayers, T. Abdul-Redah and
P.Platzman, {submitted to PRL}
\bibitem{rossky} K.F. Wong and P.Rossky, J. Chem. Phys. {\bf
116},8418, (2002)
\bibitem{bit}B. J. Schwartz, E. R. Bittner, O. V. Prezhdo and P.
Rossky, J. Chem. Phys. {\bf 104}, 5942, (1996)
\bibitem{abdul} T. Abdul-Redah, {private communication}
\bibitem{cc} P. Cabral do Couto, R.C. Guedes, and B.J. Costa
Cabral, Braz. J. Phys {\bf 34}, 42 (2004)
\bibitem{gup} M. Gupta and J. P. Burger, Phys. Rev. B {\bf 22},
6074, (1980)
\end{references}
\end{document}